\documentclass[a4paper, fleqn, 12pt]{article}

\usepackage{endfloat}
\usepackage{epsf} 
\usepackage{times}
\usepackage[latin1]{inputenc}
\usepackage{amsmath}
\usepackage{amsthm}
\usepackage{graphicx}
\usepackage{verbatim}
\usepackage[numbers]{natbib}

\newcommand{\n}{N}
\newcommand{\state}{x}
\newcommand{\State}{\varSigma}
\newcommand{\nostates}{s}
\newcommand{\statevec}{\boldsymbol{x}}
\newcommand{\transfunc}{\ensuremath{\varDelta}}
\newcommand{\nb}{\ensuremath{k}}
\newcommand{\h}{\ensuremath{\kappa}}
\newcommand{\gcaA}{\ensuremath{\Omega_1}}
\newcommand{\gcaB}{\ensuremath{\Omega_2}}
\newcommand{\gca}[1]{\ensuremath{\Omega_{#1}}}
\newcommand{\langton}{\ensuremath{\lambda}}

\newcommand{\noconfig}{\xi}
\newcommand{\rr}{DDR~}
\newcommand{\er}{ER~}
\newcommand{\W}{W} 
\newcommand{\wl}{l} 
\newcommand{\kate}{\boldsymbol{y}}
\newcommand{\E}{S} 
\newcommand{\ab}{\W\!\E}
\newcommand{\constc}{r} 

\begin{document}

\bibliographystyle{agrempe}

\begin{center}
  {\Large \bf Topology regulates pattern formation capacity of binary
    cellular automata on graphs}
  \\
  \vspace{0.5cm} Carsten Marr\footnote[2]{marr@bio.tu-darmstadt.de}
  and Marc-Thorsten H\"utt\footnote[3]{huett@bio.tu-darmstadt.de}\\
  \vspace{0.2cm} {\it Bioinformatics Group, Department of Biology,
    Darmstadt University of Technology\footnote[4]{Schnittspahnstr. 3-5,
    D-64287 Darmstadt, Germany}}
\end{center}

\begin{abstract}
  We study the effect of topology variation on the dynamic behavior of
  a system with local update rules.  We implement one-dimensional
  binary cellular automata on graphs with various topologies by
  formulating two sets of degree-dependent rules, each containing a
  single parameter. We observe that changes in graph topology induce
  transitions between different dynamic domains (Wolfram classes)
  without a formal change in the update rule. Along with topological
  variations, we study the pattern formation capacities of regular,
  random, small-world and scale-free graphs. Pattern formation
  capacity is quantified in terms of two entropy measures, which for
  standard cellular automata allow a qualitative distinction between
  the four Wolfram classes. A mean-field model explains the dynamic
  behavior of random graphs. Implications for our understanding of
  information transport through complex, network-based systems are
  discussed.
\end{abstract}

\section{Introduction}

Cellular automata (CAs) have been studied in terms of complexity
theory and computational universality and, moreover, they often serve
as models of pattern formation. The usual topologies of CAs are chains
and regular lattices. Current research on applications of graph theory
to the analysis of natural and technical systems reveals that many
real systems are based on networks with a much more complex structure
\cite{jeong00,fell00,newman01,jeong01,farkas03}. Small-world graphs
and scale-free topologies are the epitome of the huge progress in
attempts to assess observed structures and properly model
evolutionarily expanding networks
\cite{watts98,barabasi99,strogatz01}.

We, therefore, re-investigate a standard model system of
spatiotemporal pattern formation, namely binary CAs, by implementing
it on a graph and systematically varying topological features. How is
the spatiotemporal pattern of a CA (for example the capacity to
display oscillations or the complexity of an element's time evolution
within the system) changed, when, e.g., some shortcuts are introduced
into the system, or, the regular neighborhood structure is completely
substituted by a random graph topology? While it is clear that the
change of a specific CA rule alters the dynamics and, consequently,
the spatiotemporal pattern, it seems worthwhile to study the changes
under topology variation at a fixed update scheme. Some approaches
already deal with binary dynamics on complex topologies. The Ising
model for example has been implemented on a small-world graph
\cite{barrat00} and the scale-free topology has served as the backbone
for boolean dynamics \cite{aldana03}, as well as for the SIS and SIR
epidemic models \cite{pastor-satorras01,barthelemy04}. Only few
attempts investigate the connection between changes in topology and
the corresponding impact on CA dynamics.  In \cite{huang03} the ``game
of life'' is studied on a small-world network and a phase transition
at a critical network disorder is found.  However, the link between
dynamic features and topology variation, especially for
one-dimensional CAs, has not been addressed yet.  Therefore, in this
paper we pursue the question, to what extent topology determines and
constrains the capacity of a
system to display certain forms of dynamics.\\

In the following, we briefly review the cellular automaton (CA)
formalism and introduce the notation used throughout the paper.

A CA describes the deterministic time evolution of regularly coupled
cells. Formally, a one-dimensional (1D) CA consists of a chain of $\n$
cells in states $\state \in \State$, with the set of all possible cell
states $\State$, and a transition function $\transfunc$. The standard
topology of a 1D CA is a ring lattice where node $i$ is connected to
its $\nb$ next neighbors ranging from $i-\nb/2$ to $i+\nb/2$ with even
$\nb$. The transition function $\transfunc$ maps the configuration
$\tilde{\kate}_i$ of cell state $i$ together with its $\nb$
neighboring cell states at time $t$, onto the state of the central
cell $i$ in the next time step, $\transfunc: \tilde{\kate}_i(t) = \{
\state_{i-\nb/2}(t), \ldots , \state_i(t), \ldots, \state_{i+\nb/2}(t)
\} \rightarrow \state_i (t+1)$. For $\nostates =|\State|$ different
cell states there exist $\noconfig = \nostates^{\nb+1}$ different
neighborhood configurations.  Each transition function $\transfunc$
maps these configurations on elements in $\State$, resulting in
$\nostates^\noconfig$ possible transition functions. This
combinatorial aspect of the set of possible transition functions is
also often used to attribute rule numbers to the $\nostates^\noconfig$
update rules. To this end, the neighborhood configurations
$\tilde{\kate}$ are ordered according to the increasing value of the
corresponding binary number. In that way sorted, the list of mapped
states represents the rule number in binary digits.  In a system-wide
picture, $\transfunc$ defines the transitions between the
$\nostates^{\n}$ different system configurations $\statevec = \{
\state_1, \state_2, \ldots , \state_n \}$. From this point of view,
the configuration space together with the update rule $\transfunc$
forms a directed network, where the transitions between the system's
configurations are completely determined. After at most
$\nostates^{\n}$ time steps, any finite-size CA consisting of $\n$
elements revisits a previously encountered state and repeats its
dynamics. This network-like aspect of CAs in configuration space is
frequently discussed in studies of boolean dynamics.

We will restrict our investigation throughout this paper to a binary
state space $\State=\{0,1\}$, and update all cells synchronously at
every time step. The minimal dimension $\nostates =2$ of the state
space keeps the discussion of dynamic properties simple
and, furthermore, allows for parallels to previous dynamic models and work on
elementary CAs \cite{wolfram83}.

Different attempts to classify the rule space of CAs have been carried
out.  Wolfram \cite{wolfram84} divided CAs qualitatively into four
classes, according to the emerging spatiotemporal patterns and
analogous to dynamical systems descriptions: I (homogeneous stationary
state), II (heterogeneous stationary state or simple periodic
structures), III (chaotic behavior), and IV (long range correlations
and propagating structures). The introduction of the Langton parameter
$\langton$ \cite{Langton90} allowed a quantitative investigation of CA
rules, even for large rule spaces. While for $s>2$ Langton's scheme of
generating a transition function for a given value of $\langton$
requires some statistical subsidiary conditions, for binary CAs
$\langton$ is simply the number of neighborhoods mapped onto the state
$1$ divided by the number of all possible neighborhood states. For
rules generated with the ``random-table method'', where $\langton$ is
used as the probability for a neighborhood to be mapped on state 1,
the Langton parameter defines trajectories through the CA rule space
and, consequently, through the four different dynamical
regimes. The order, in which the corresponding Wolfram classes are
passed as $\langton$ is increased, is I $\to $ II $\to $ IV $\to $
III.\@  Class IV automata, lying between periodic (II) and chaotic
(III) behavior and exhibiting long-range correlations and propagating
structures, are regarded as suitable scenarios for the study of
complexity and self-organization. A recent, alternative parametric
approach is given in \cite{sakai04}.  There, the authors introduce a
parameter $F$ controlling details of the transition function beyond
the $\langton$ parameter and show the existence of all Wolfram classes
for a given $\langton$ and an appropriate $F$. Obviously, the
diversity of classes in a subset of rules with a given $\langton$
depends on the way the rules are generated.\\

Our paper is structured as follows: In Section 2, we formulate two
classes of binary cellular automata on graphs and discuss the
resulting dynamics. While the first class keeps the Langton parameter
$\lambda$ constant, $\lambda $ varies with the parameter $\h$ for the
second class.  In the Section 3, we focus on complex spatiotemporal
patterns as an indicator of optimal information transport and
introduce statistical tools for segregating different dynamic domains.
Applied to ordinary CAs, the Shannon entropy $\E$ and the word entropy
$\W$ allow an adequate qualitative separation of the four Wolfram
classes.  Similarly to Langton's investigation, the pattern formation
capacity of cellular automata can thus be quantitatively analyzed in a
two-dimensional plane spanned by the two entropy measures.  In
Section 4 we continuously change topological parameters of the
conventional ring lattice and implement the two sets of binary
dynamics.  A link between CAs on graphs, as described in this
investigation, and more traditional forms of CA studies is provided by
analyzing regular graph topologies with different neighborhood sizes.
We then study procedures of topology variations and compare the
emerging patterns of small-world and random graphs with the
well-classified ones from CAs in terms of entropy measures. Finally,
we study the pattern formation capacity of scale-free graphs and
randomized variants with the same degree distribution.
Section 5 gives a mean-field analysis of the dynamic behavior of
random graphs based on the local state densities which govern the
nodes' dynamics. In this way we gain insight into the mechanisms of the
considered dynamics and the link between some topological and
dynamical features.
The last section (Section 6) reviews the results, discusses
implications for the dynamics on real networks and lines out some
ideas for further investigations.

\section{Cellular automata on graphs}

The topology of a 1D CA is usually a regular ring lattice with a
clustered neighborhood structure, i.e.\ a graph of $\n$ nodes forming
a closed chain with additional edges linking each element to a certain
number of neighbors.  In such a graph each node has the same degree
$\nb$ due to the links to the $\nb/2$ neighbors in both directions
along the chain.  In this paper we will refer to this specific ring
lattice topology as a regular graph. Note, however, that in graph
theory this term is used for all networks with a delta-like degree
distribution.  Since in CAs the transition function $\transfunc$ is
defined for a fixed number of neighbors $\nb$, this system does not
provide an appropriate framework for studying changes in the system's
architecture. A change of $\nb$ would correspond to an alteration of
the underlying rule space, which restricts comparability of two
systems with different degree. We therefore introduce two sets of
functions $\gca1(\h)$ and $\gca2(\h)$, each depending on a single parameter $\h$,
which account for the individually varying architecture and,
consequently, allow the formulation of binary CAs on arbitrary graphs.

Let the graph be represented by its adjacency matrix $A_{i\!j}$. When
a link exists between nodes $i$ and $j$ of the graph, the
corresponding matrix element is 1, and it is 0 otherwise. We consider
only undirected graphs. Consequently, the adjacency matrix is
symmetric, $A_{i\!j} = A_{j\!i}$. We call the first set of rules
\gca1. There, the state $\state_i$ of node $i$ of a graph flips its
state in the next time step if the density of 1's among the $\nb_i$
elements linked to node $i$, $\rho_i$, is larger than a parameter
$\h$,
\begin{equation}
  \label{eq:spieldef1}
  \gcaA(\h): \; \state_i(t+1) = 
  \begin{cases}
    \state_i(t) \,,    &  \rho_i \leq \h \\[0.2cm]
    1- \state_i(t) \,,  &  \rho_i > \h \;.
    \end{cases}
\end{equation}  
There, $\rho_i$ is defined by 
\begin{equation}
  \label{eq:r1}
  \rho_i = \frac{1}{\nb_i} \sum_{j} A_{i\!j} \state_j(t) \;. 
\end{equation}
We will refer to this quantity as the local density. Since the sum does not include $\state_i$
itself, the action on node $i$ is invariant under the changing of the
central element.  This symmetry is broken by introducing
\begin{equation}
  \label{eq:spieldef2}
  \gcaB(\h): \; \state_i(t+1) = 
  \begin{cases}
    \state_i(t) \,,    &   \tilde{\rho}_i \leq \h \\[2mm]
    1-\state_i(t) \,,  &   \tilde{\rho}_i > \h \; ,
    \end{cases}
\end{equation}  
where now the density $\tilde{\rho}_i$ incorporates the state $\state_i$
itself: 
\begin{equation}
  \label{eq:r2}
  \tilde{\rho}_i = \frac{1}{\nb_i + 1} \left( \state_i(t) + \sum_j A_{i\!j} \state_j(t) \right) \; .
\end{equation}

In $\gca{1}$ the state $\state_i(t+1)$ depends on the sum
over the neighboring states and on the state $\state_i(t)$ itself,
\begin{equation}
  \label{eq:dahl}
  \gca{1}: \; \state_i(t+1) = f \left(  \state_i(t), \sum_j A_{i\!j}
  \state_j(t) \right) \; , 
\end{equation}
whereas in $\gca{2}$ the state $\state_i(t)$ is also included in the sum.
Notably, all rules defined by \gca{i} are legal, according to
Wolfram's definition \cite{wolfram83}, i.e.\ they fulfill the quiescent
condition (a configuration of only zeroes is unchanged) and reflection
symmetry (001 and 100 are mapped onto the same state). \\

In order to get familiar with the set of functions $\gcaA$ and
$\gcaB$, it is instructive to apply the rules to regular graphs first.
In this case, each $\gca{i}(\h)$ is identical to a corresponding CA
rule \transfunc. Table~\ref{tab:rules} shows the neighborhood mappings
for an elementary CA, i.e., a CA with $\nostates=2$ and $\nb = 2$. By
inspecting the rule table, we can easily infer the corresponding CA
rules: $\gca{1}(0 \leq \h < \frac{1}{2}) \equiv \textrm{rule} \;
00110110_2 \equiv \textrm{rule} \; 54$ etc. Earlier investigations on
elementary CA \cite{wolfram83} showed that these rules correspond to
stationary (rules 36, 76, 108 and 204), oscillatory (rule 50) and
complex (rule 54) behavior, which can also be inferred in a natural
way from the definitions (\ref{eq:spieldef1}) and
(\ref{eq:spieldef2}): According to (\ref{eq:spieldef1}), for small
$\h$ most nodes will change their state and an oscillatory behavior
will dominate, while for large $\h$ the majority of nodes will not change
their state and a stationary pattern will prevail. In between these
limiting regimes we expect complex and chaotic behavior.

\begin{table}[h]
\centering
\begin{tabular}[h]{c|ccc|cccc}
  \multicolumn{1}{c}{}&
  \multicolumn{3}{c}{\gca{1}}&
  \multicolumn{4}{c}{\gca{2}}\\[0.2cm]
    $\nb$ &
    $0 \le \h < \frac{1}{2}$  & $\frac{1}{2} \le \h < 1$ & $\h=1$ &
    $0 \le \h < \frac{1}{3}$ &  $\frac{1}{3} \le \h < \frac{2}{3}$ & 
    $\frac{2}{3} \le \h < 1$ & $\h=1$  \\[0.1cm]
    \hline
    &&&&&&&\\[-0.2cm]
    000 & 
    0 & 0 & 0 &
    0 & 0 & 0 & 0 \\
001 &
{\bf 1} & 0 & 0 &
{\bf 1} & 0 & 0 & 0  \\
010&
1&1&1&
\bf0&1&1&1 \\
011&
\bf0&1&1&
\bf0&\bf0&1&1\\
100&
\bf1&0&0&
\bf1&0&0&0\\
101&
{\bf1}&{\bf1}&0&
{\bf1}&{\bf1}&0&0\\
110&
\bf0&1&1&
\bf0&\bf0&1&1\\
111&
\bf0&\bf0&1&
\bf0&\bf0&\bf0&1\\[0.1cm]
\hline
&&&&&&&\\[-0.3cm]
\transfunc & 
54 & 108 & 204 & 
50 & 36 & 76 & 204
\end{tabular}
\caption{Neighborhood mappings of rules $\gca{1}$ and $\gca{2}$ for
  elementary CAs, i.e.\ with $\nb=\nostates=2$. The number of different rules is
  $\nb$ and $\nb+1$ for $\gca{1}$ and $\gca{2}$, respectively. If the
  state of the node changes, the corresponding entry is shown in bold
  face. In the last row, the corresponding CA rule numbers are shown.}
  \label{tab:rules}
\end{table}

As can be inferred from Tab.~\ref{tab:rules}, the Langton parameter of
all $\gcaA$ rules is constant, $\langton=0.5$, since the number of
neighborhoods mapped on 0 and 1 is equal. This is a generic property
of $\gca{1}$. Let $\kate_i = \{x_{i-k/2}, \ldots, x_{i+k/2} \}$ denote
the neighborhood configuration of node $i$ excluding the state $x_i$
itself.  Then, under $\gca1(\h)$ a neighborhood configuration $\kate$
will flip a central 0 as well as a central 1 for an appropriate value
of $\h$. Since the identical mapping (rule 204) is characterized by
$\langton=0.5$ and the number of flipping 0's and 1's is equal,
$\langton = 0.5$ for all \gca1 rules.  Therefore, this set of
rules exploits a pattern formation capacity orthogonal to Langton's
investigation and, as we will see, proves the existence of all dynamic
domains for $\langton=0.5$. This is an elementary example of changing
the Wolfram class at constant $\langton$ as stressed in
\cite{sakai04}.

The lack of the invariance with respect to the central element in $\gcaB$ is
responsible for a varying Langton parameter in this case.  We find
that the number of 1's in the rule table (and accordingly $\langton$)
varies with $\h$, as shown in Tab.~\ref{tab:rules}. In the regime
where oscillatory and stationary behaviors coexist and complex
behavior can be expected (i.e.\ $\frac{1}{3} \le \h < \frac{2}{3}$)
most neighborhoods are mapped on 0 and the system is prone to a
density loss in the course of time.

\section{Quantitative classification of spatiotemporal patterns}

In this section we present two entropy measures and apply them to
conventional CAs in order to separate and categorize general spatiotemporal
patterns of such CAs. By selecting only quantities evaluating the time
development of single elements (as opposed to whole spatiotemporal
configurations) this classification scheme can, after gauging it with
conventional CAs, be directly applied to CAs on graphs, where the
degree $\nb$ is given by a distribution $P(k)$.

The Shannon entropy $\E$ relies on the probability for the emergence
of the cell states $\state \in \State$ in the time evolution of cell $i$,
\begin{equation}
  \label{eq:4}
  \E_i = - \sum_{j=0}^{\nostates - 1} p(\state_j) \log_2 p(\state_j) \; .
\end{equation}
For binary dynamics, the $p(\state_j)$ are the probabilities for 0's
and 1's. Averaging $\E_i$ over all $\n$ cells yields the average
Shannon entropy $\E = \sum_{i=1}^\n \E_i/\n$.  All stationary patterns
result in $\E=0$, while random and oscillatory patterns yield $\E=1$
because of the equal distribution of 0's and 1's in the time
series of each cell.

The word entropy $\W$ uses the occurrence of blocks of constant cells
of length $\wl$ ($\wl$-words) in the time series of cell $i$,
\begin{equation}
  \label{eq:6}
  \W_i = - \sum_{\wl=1}^{t} p(\wl) \log_2 p(\wl) \; ,
\end{equation}
where $p(\wl)$ is the probability for an $\wl$-word, irrespective of the
state $\state$ this word consists of. The maximal word length is clearly the
length of the time series $T$. The average value of the word entropy
is given by $\W = \sum_{i=1}^\n \W_i/\n$.  The word entropy is similar
to Wolfram's ``measure entropy'' \cite{wolfram84}, but instead of
spatial blocks we use temporal correlations here. Moreover, $\W$
measures solely the occurrence of blocks of constant cells and not all
possible blocks of length $\wl$.

Both entropy measures analyze the temporal behavior of cells. The
drawback of this definition is clearly the dependence of $\W$ on the
length $T$ of the time series of the pattern. However, this problem
exists for the spatial variant (i.e.\ the measure entropy), when
finite-size CAs are considered. As pointed out above, the advantage of
the measures presented here is their generic applicability to
arbitrary topologies, where the analyzed spatiotemporal patterns lose
the spatial information because of the irregular neighborhood
configurations.

Alternatives for the word entropy are given by, e.g., the mutual
information and by spectral properties of average-density
fluctuations. The Shannon entropy on the other hand exhibits partially
similar properties as the Hamming distance. We checked that all these
quantities lead to analog qualitative results.\\

\begin{figure}[h]
  \centering
  \epsfxsize=14cm
  \epsffile{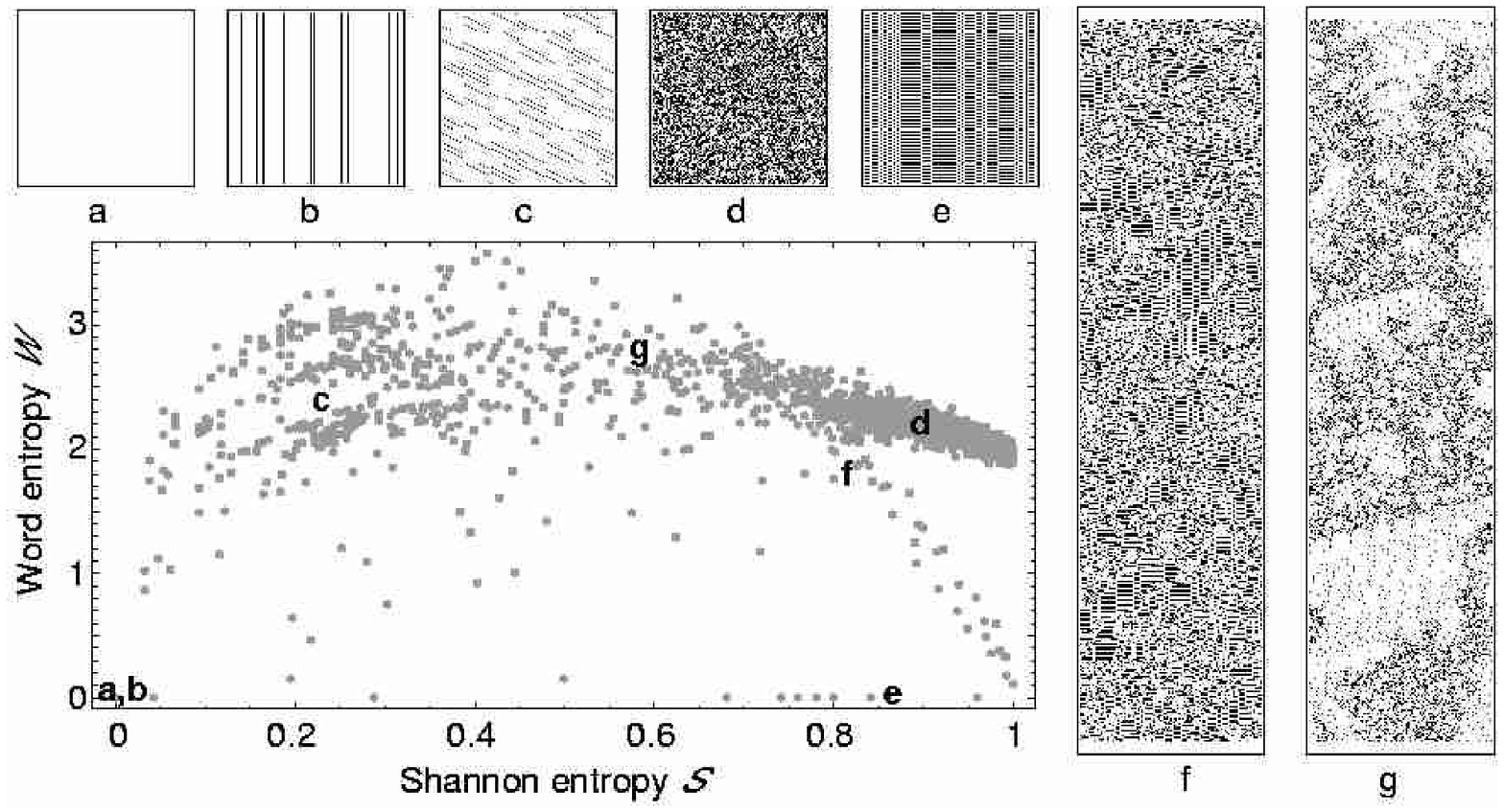}
  \caption{Typical patterns of the different dynamic domains
    within the $\ab$ plane. The plane is filled with $\n=500$,
    $\nb=10$ CAs where the time interval $]500,1000]$ has been used to
    calculate $\W$ and $\E$. Patterns of stationary, oscillatory,
    periodic and chaotic automata, (a) - (f), comprise 500 time steps,
    the two class IV patterns (f) and (g) comprise 2000 time steps.
    0's are indicated black, 1's white.}
  \label{klassen}
\end{figure}

With these tools available we can visualize the separation of CAs
within a $\ab$ plane. Rather than inferring rule space properties,
we want to locate domains of spatiotemporal patterns on this plane. We
therefore gauge the plane with $\nb=10$ CAs, where the initial cell
states 0 and 1 have been randomly assigned. We generate the different
CA rules according to Langton's ``random-table method'', but other
parameterizations like the one proposed in \cite{sakai04} could be used as well.
Fig.~\ref{klassen} shows typical patterns of $\n=500$, $\nb=10$ CAs.
We can now classify the patterns according to their $\W$ and $\E$
values within the $\ab$ plane. Stationary patterns lie in the lower
left corner of the plane, while purely oscillatory ones are located in
the lower right. Chaotic patterns are localized in a rather
small region in the upper right part of the plane. Between these 
extremal regions, the plane is filled with CAs with partly non-trivial
periodic structures and patterns with long-range correlations.

We can now plot the $\gca{1}(\h)$ and $\gca{2}(\h)$ rules in the $\ab$
plane to visually infer the effect of increasing $\h$.  All the
following simulations have been carried out on graphs with 500 nodes
and random initial condition, i.e., every cell is independently
assigned a value $\state_i \in \{0,1\}$ at time $t=0$, resulting in an
initial density $\rho(t=0)$ of about $0.5$. We always drop the first
500 time steps and use the second 500 steps to calculate the entropy
measures. We display the medians of five runs in all following figures
to ensure statistical reliability.  

\begin{figure}[h]
  \centering
  \epsfxsize=11cm
  \epsffile{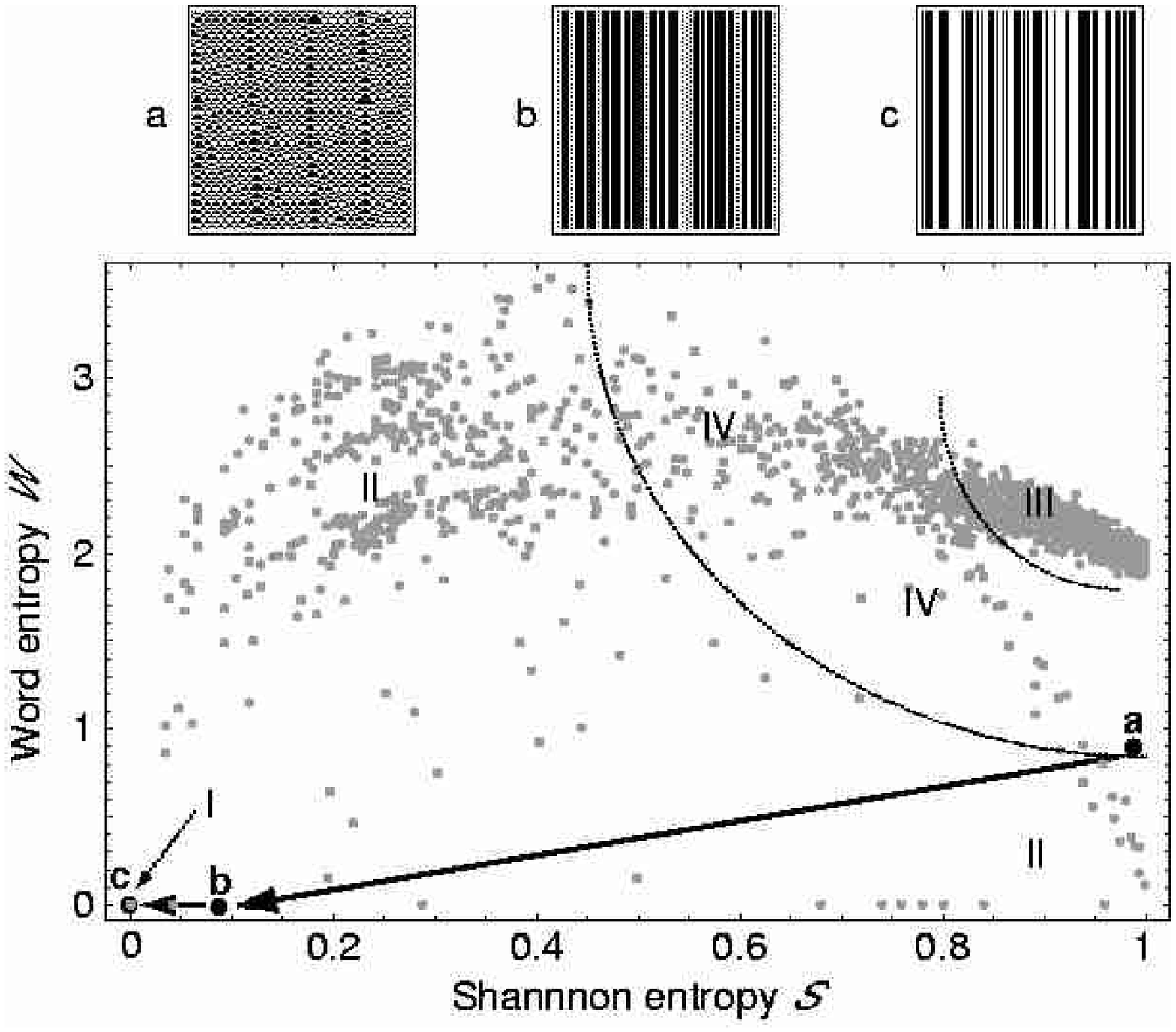}
  \caption{Behavior of $\gca{1}$ for elementary, i.e.\ $\nb =
    \nostates = 2$, CAs within the $\ab$ plane as $\h$ is increased.
    The three possible neighborhood configurations correspond to the
    rules 54, 108, 204 (see Tab.~\ref{tab:rules}). For computation of
    $\W$ and $\E$ here and in the following the same system size as in
    Fig.~\ref{klassen} has been used ($N=500$, $T=1000$) with a
    transient of $500$ time steps being dropped. The patterns
    represent the three possible domains and show a selection of 100
    nodes for 100 time steps for visual clarity. Note that there are
    no precise borders between the different regions in the $\ab$
    plane.  Therefore the highlighted regions reflect a tendency
    rather than a strict distinction between two adjacent Wolfram
    classes.}
  \label{fig:lasagne}
\end{figure}

In Fig.~\ref{fig:lasagne} we show the gauged plane with the four
Wolfram classes qualitatively assigned to specific regions. This
allows the immediate identification of the different dynamic domains
as $\h$ is increased for $\gca{1}(\h)$ on a regular graph with 2
neighbors or, in the language of CAs, the succession of rules 54, 108,
204 for $\nb=2$ CAs. Three snapshots show the corresponding patterns,
where we selected 100 nodes and 100 time steps for visual clarity.

\begin{figure}[h]
  \centering
  \epsfxsize=11cm
  \epsffile{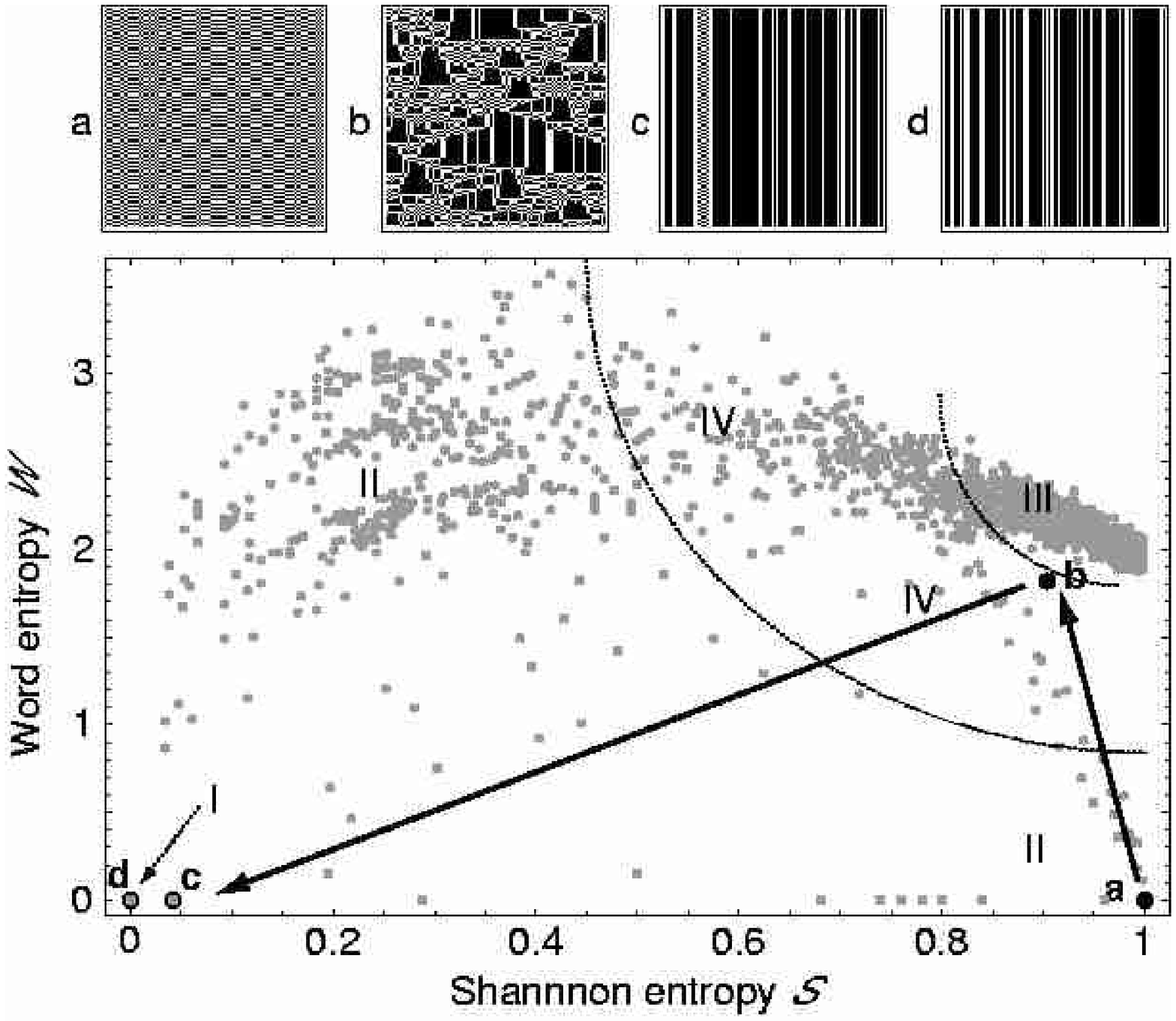}
  \caption{Behavior of $\gca{1}$ for $\nb=10$ CAs within the $\ab$ plane as $\h$ is
    increased. Only for $\h \in [0.3,0.4[$ (b) complex patterns
    emerge, for all other parameters, the corresponding rules result
    in purely oscillatory and stationary patterns respectively. The other
    spatiotemporal patterns shown in the top row correspond to
    parameter values $\h \in 
    [0.2,0.3[$ (a), $\h \in [0.4, 0.5[$ (c) and $\h \in [0.5, 0.6[$
    (d), again for 100 nodes and 100 time steps.}
  \label{fig:fleks}
\end{figure}

A larger number of rules with Langton parameter $\langton=0.5$ can be
studied for a larger neighborhood size. Fig.~\ref{fig:fleks} shows the
trajectory of $\nb=10$ CAs for increasing $\h$ in the gauged $\ab$
plane. In this case, the transition from oscillatory to stationary
behavior is accompanied by complex structures in the region of class
IV automata. Large $\W$ values emerge for every $\h \in [0.3,0.4[$,
where the exact parameter value is irrelevant because of the discrete
number of different local densities. 

\begin{figure}[h]
  \centering
  \epsfxsize=11cm
  \epsffile{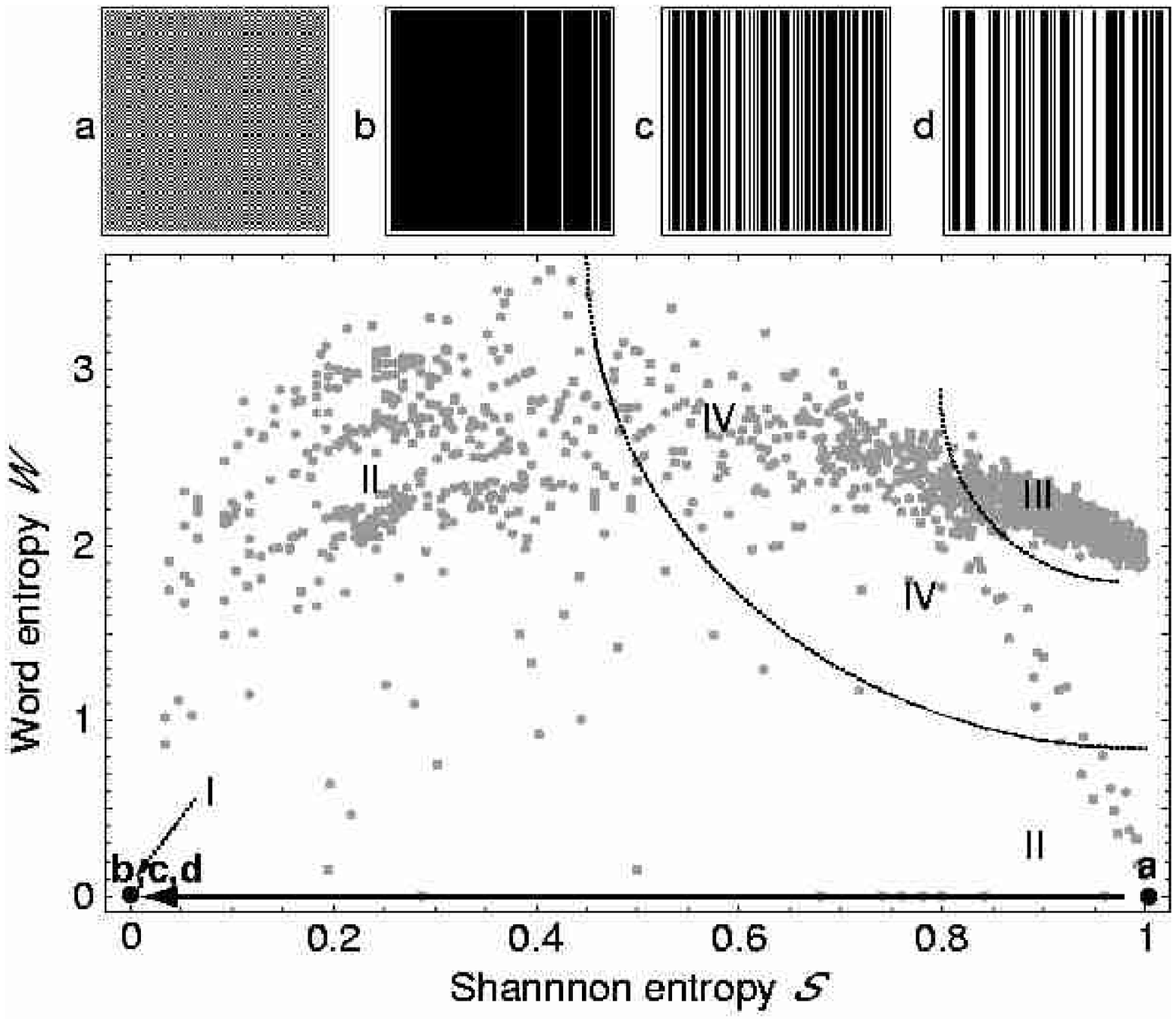}
  \caption{Behavior of \gca{2} for elementary CAs within the $\ab$ plane as $\h$ is
    increased. The corresponding rule numbers for the elementary CA
    are 50, 36, 76, 204 (see Tab.~\ref{tab:rules}). Again, the
    patterns show 100 nodes for 100 time steps. All stationary patterns
    are mapped onto the same point in the $\ab$ plane.}
  \label{fig:lasagne2}
\end{figure}

Fig.~\ref{fig:lasagne2} shows $\gca{2}(\h)$ for $\nb = 2$ CAs. In
the region $\frac{1}{3} \le \h < \frac{2}{3}$, where one would expect
complex patterns analogous to $\gca{1}$, only two neighborhood
configurations are mapped onto the state 1 (see Tab.~\ref{tab:rules}),
resulting in a density loss and a stationary pattern where 0's prevail
(Fig.~\ref{fig:lasagne2}b).

\section{Topology variation}

For \gca1 the variation of $\h$ for regular graphs explores the rule space
orthogonal to the Langton parameter. Different neighborhood sizes
result in different trajectories within the $\ab$ plane. We find the
Wolfram classes II,IV and I as $\h$ is increased. This transition from
oscillatory to stationary behavior is a generic property of the sets
$\gca{i}$.

\begin{figure}[h]
  \centering
  \epsfxsize=10cm
  \epsffile{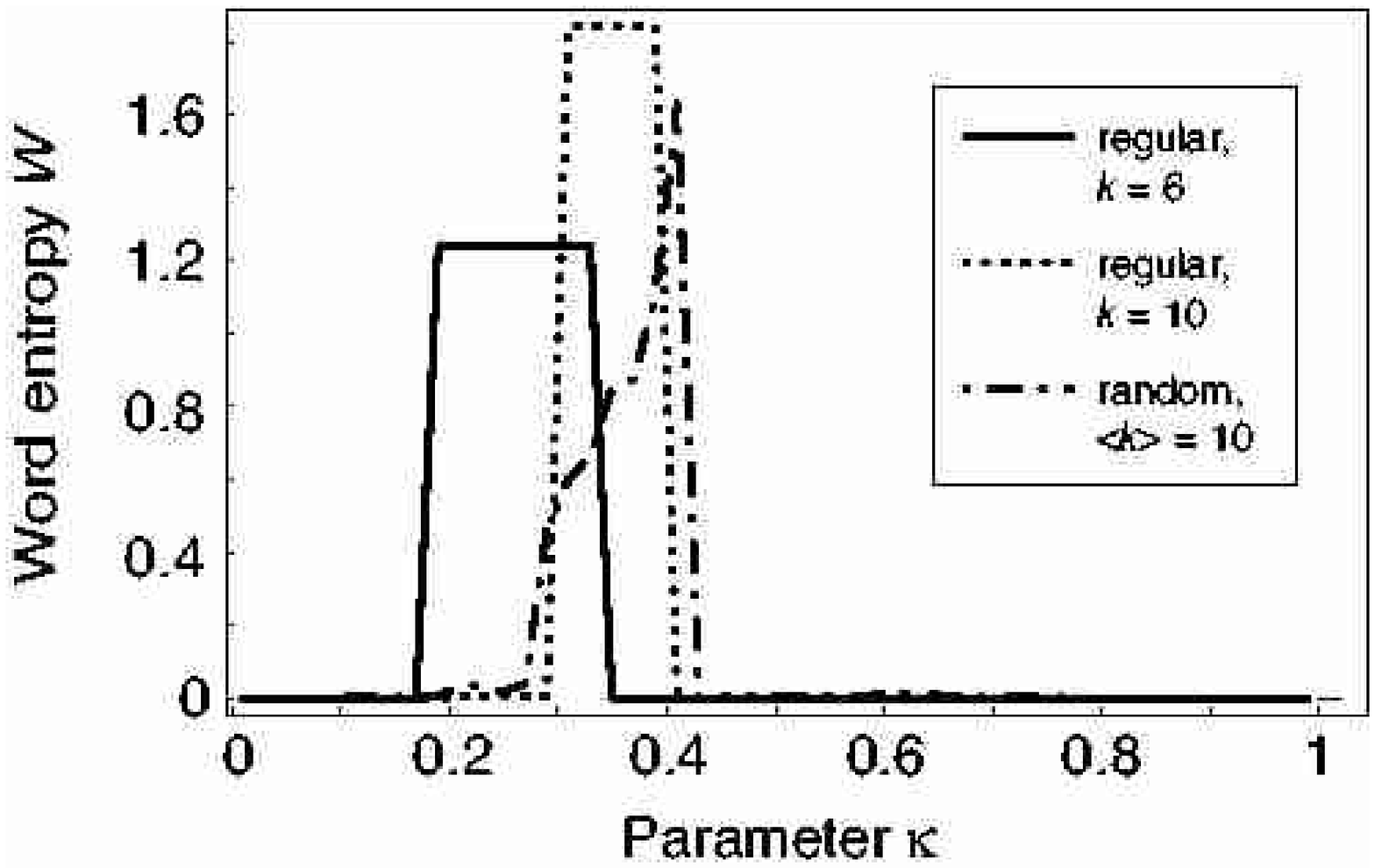}
  \caption{The word entropy $\W$ against the variation of $\h$ for
  $\gca{1}$ for two regular graphs and an $\er$ random graph. Complex
  patterns prevail in the interval between $0.3< \h < 0.4$ for the two
  graph types.}
  \label{fig:nutella}
\end{figure}

However, in this chapter we want to focus on another feature of the
two rule sets, namely their applicability to arbitrary topologies.
Regardless of the global network structure, single nodes evolve
according to their local density in every time step. We can
monitor the dynamical changes of a network along the topological
variations. To do so, we first have to select an adequate $\h$ value
where the capacity of producing complex patterns is high, as we expect
that in this configuration the patterns are most sensitive to the topology
of the underlying graph. Fig.~\ref{fig:nutella} shows the word entropy
vs. $\h$ for $k=6$ and $k=10$ regular graphs and, since we will deal
with random graphs later on, for a random graph of the Erd\H{o}s and
Rényi type \cite{erdos59} with a mean degree of $10$.
In order to encounter a large variety of patterns and high word
entropies for both regular and random graphs, we choose a $\h$ value
within the interval $[0.35, 0.4[$. In the following simulations,
$\h=0.36$ will be used as the parameter value for the $\gca{i}$, but
similar values of $\h$ would serve as well, as can be inferred from
Fig.~\ref{fig:nutella}.\\

\begin{figure}[h]
  \centering
  \epsfxsize=9cm
  \epsffile{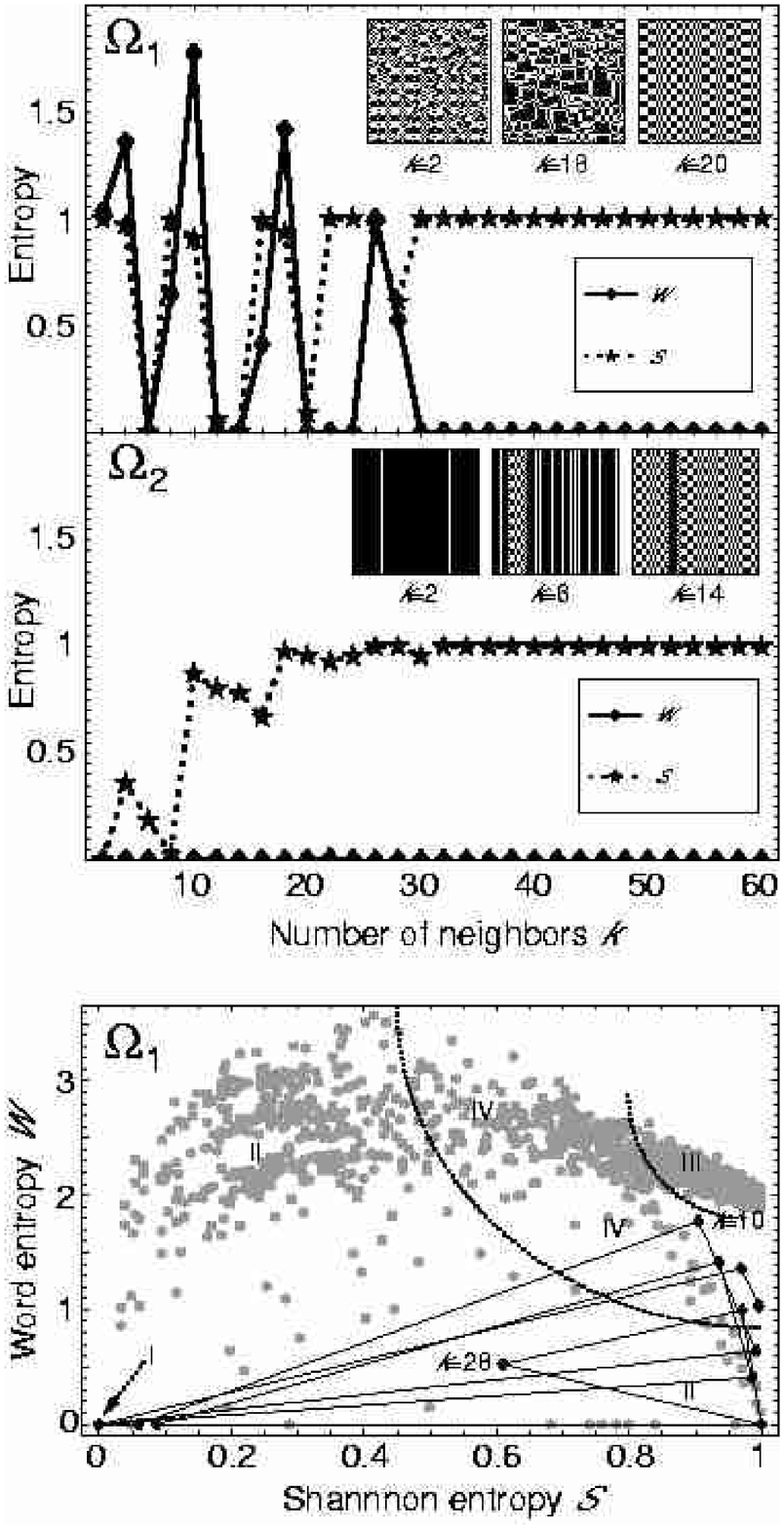}
  \caption{Word entropy (solid) and Shannon entropy (dashed) as a
    function of neighborhood size $\nb$ for $\gca{1}$ and $\gca{2}$
    (top). The inset patterns show the evolution of 100 nodes for 100
    time steps for three different values of $\nb$. The lower picture
    shows the trajectory of regular graphs in the \ab plane under
    \gca1 as $\nb$ is increased.}
  \label{fig:sizi}
\end{figure}

As a first step we continue analyzing regular graphs --- and thereby
allow for conventional CA interpretation --- but increase the
neighborhood size $\nb$ for every node in even numbers.  In this case,
$\gca{i}(0.36)$ corresponds to a different CA transition rule
$\transfunc$ for each neighborhood size \nb.  Fig.~\ref{fig:sizi}
shows $\W$ and $\E$ of spatiotemporal patterns of $\gca{1}$ and
$\gca{2}$ (top) on regular graphs against increasing neighborhood size
$\nb$. The small difference in the definition of the two sets of rules
leads to striking differences of the dynamic response for regular
graphs. For small $\nb$ we find a multiple peak structure in the
entropy measures for $\gca{1}$ and a variety of different
spatiotemporal patterns. For $\gca{2}$, the word entropy $\W$ is close
to 0 for all neighborhood sizes. For large neighborhoods around $\nb >
30$, the dynamics is purely oscillatory, indicated by $\E=1$ and
$\W=0$. The lower picture in Fig.~\ref{fig:nutella} shows the path for
\gca1 in the $\ab$ plane as $\nb$ is increased. The peak structure in
the picture above appears now as jumps between different Wolfram
classes.

The entropy values for large $\nb$ can be understood in the limiting
case of a completely connected graph (complete graph) which
corresponds to CAs with $\nb=\n-1$.  There, the neighborhood is
approximately identical for all nodes and the overall dynamics is
governed by the initial density $\rho(t=0)$. For $\rho(0) \le \h$ all
nodes remain in their initial state, for $\rho(0) \land 1-\rho(0) >
\h$ the nodes oscillate constantly, while for $1-\rho(0) \le \h <
\rho(0)$, all nodes flip their state in the first time step and then
remain constant. Above a certain critical value of the connectivity,
all elements have essentially the same neighborhood and consequently
display the same dynamical behavior.  This threshold is independent of
other topological details and corresponds to the synchronization
threshold known from (continuous) dynamical systems.

If we decrease $\nb$, the densities of 1's in the neighborhoods
$\kate_i = \{x_{-k/2}, \ldots, x_{k/2} \}$ increasingly differ and at
some point, these differences lead to a coexistence of stationary and
oscillatory behavior, i.e., complex system dynamics can emerge. If we
regard $\gca{2}$, we find that the coexistence of stationary and
oscillatory behavior is not sufficient for the emergence of complex
system dynamics. There, the transition between the stationary and
oscillatory domains is accompanied by spatially clearly separated
blocks of different behavior, as can be seen in the pattern examples
in Fig.~\ref{fig:sizi}.\\

\begin{figure}[h]
  \centering
  \epsfxsize=14cm
  \epsffile{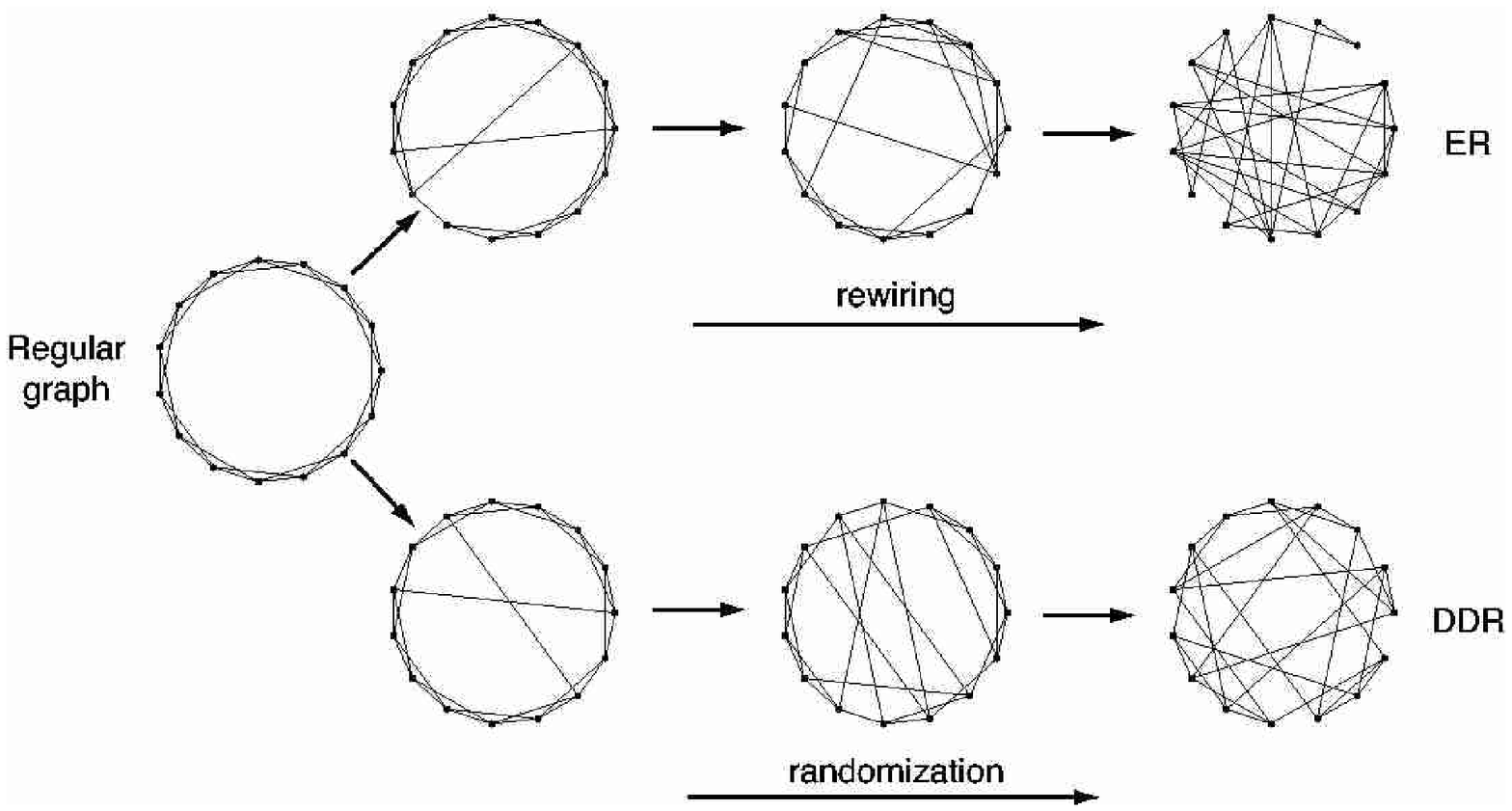}
  \caption{Rewiring (top) and randomization (bottom) of a regular
    graph. In the first case, the resulting degree distribution
    $P(\nb)$ is binomial and nodes with $\nb=2$ occur, while in the
    second case, the original degree distribution $P(k)=\delta_{k,4}$
    is conserved during the process.}
  \label{fig:rewiring}
\end{figure}

A second way to study the influence of topology is to gradually rewire
the original regular graph. We can conduct this procedure with the
preservation of the degree distribution $P(k)$ (randomization) or
without preserving $P(k)$ (rewiring), resulting in two different
topologies. If the degree distribution $P(k')=\delta_{\nb,k'}$ is
conserved, we end up with a regular graph where the $\nb$ couplings of
every node are randomly chosen out of the $\n-1$ remaining nodes. We
will refer to this type of graph as the \rr (delta-distributed random)
graph. Note that \rr graphs can also be considered as undirected
Kauffman networks \cite{kauffman69}. In the other case, where the degree
distribution is altered in the course of the rewiring, we end up with
a binomial degree distribution and a mean degree $\nb$. This limiting
case coincides with the \er graph discussed above. Such a rewiring
procedure has first been introduced by Watts and Strogatz
\cite{watts98} as a model of small-world graphs.
Fig.~\ref{fig:rewiring} visualizes the two methods for a regular ring lattice
with $N=15$, $\nb=4$.

\begin{figure}[h]
  \centering
  \epsfxsize=10cm
  \epsffile{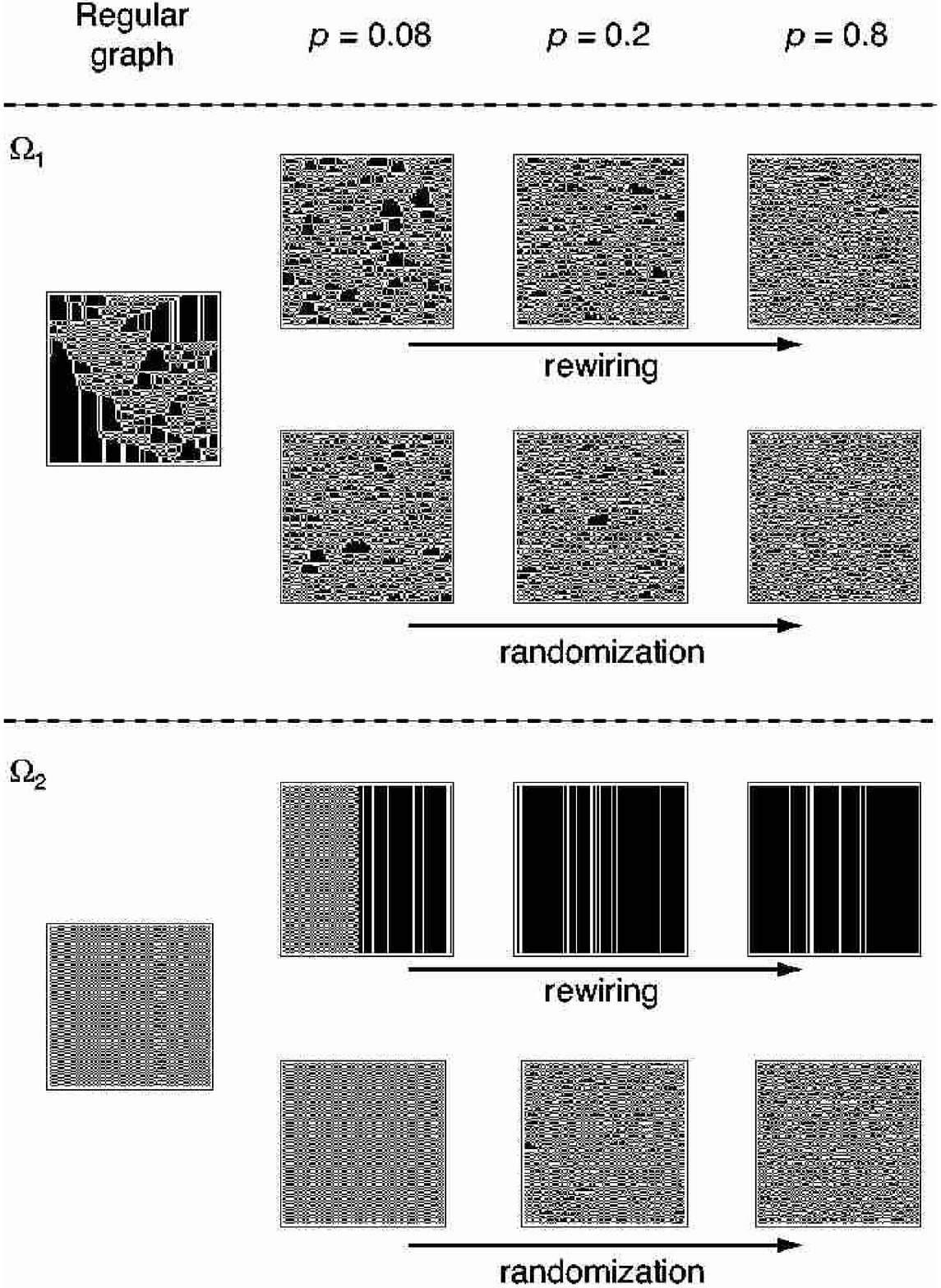}
  \caption{Spatiotemporal patterns for the different rewiring
  mechanisms described above for increasing rewiring and randomization
  depth $p$. The resulting patterns differ
  considerably for \gca1 and \gca2.} 
  \label{fig:top2_patterns}
\end{figure}

In both cases, the strict analogy to a CA and a concrete rule
numbering is lost as soon as the architecture differs from a regular
nearest-neighbor configuration and the applicability of the $\gca{i}$
to arbitrary topologies is exploited. In Fig.~\ref{fig:top2_patterns}
we display the emerging spatiotemporal patterns as the rewiring and
randomization depth $p$ is increased. This quantity is the ratio of
rewired links to all existing links. While for $\gca1(\h=0.36)$, both
procedures result in similar patterns, \gca2 resolves the different
degree distributions: the two architectures (\er and \rr) lead to
stationary and non-trivial patterns respectively. However, we have to
note that the linear arrangement of nodes according to the node number
in the space-time plots has no topological foundation. Adjacent nodes
in the pattern do no longer have to be linked to each other as the
underlying graph is rewired.

\begin{figure}[h]
  \centering
  \epsfxsize=14cm
  \epsffile{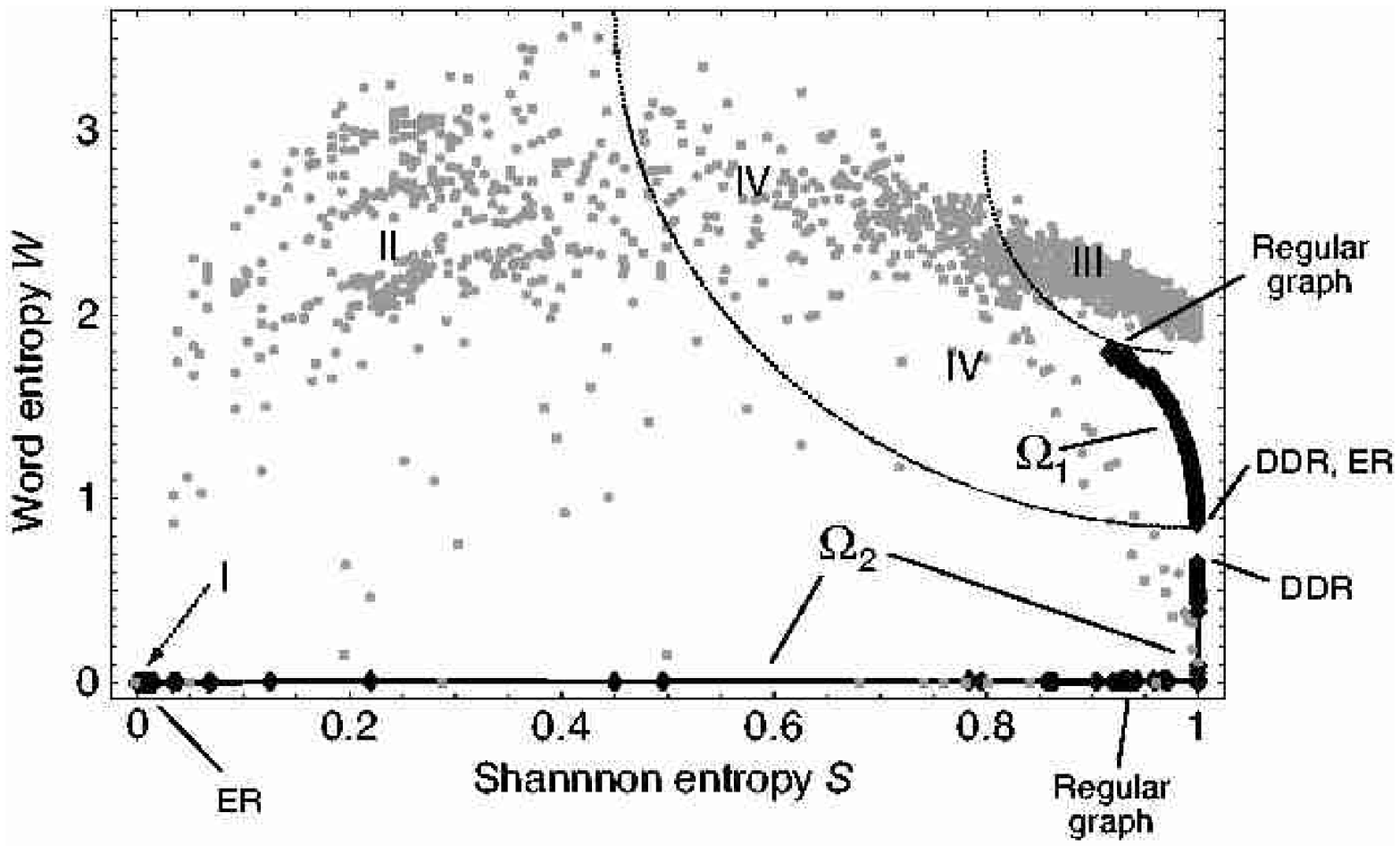}
  \caption{The four trajectories for \gca1 and \gca2 for the rewiring
  and randomization process of an originally regular graph with
  $N=500$, $\nb=10$. The two paths for \gca1 are nearly identical
  while the ones for \gca2 differ enormously.} 
  \label{fig:top2_trajectory}
\end{figure}

Fig.~\ref{fig:top2_trajectory} visualizes the trajectories of the two
procedures for \gca1 and \gca2 in the gauged $\ab$ plane and locates
the corresponding pattern formation capacity. The decrease of the word
entropy for \gca1 relies on the continuous disintegration of clustered
neighborhoods. For \gca2 we find the opposite behavior: The clustering
inhibits complex patterns for this parameter setting and only the
randomization of links allows for non-trivial structures. The large
coverage of dynamic domains demonstrates the large and systematic
regulation of pattern formation capacity by topology.

\begin{figure}[h]
  \centering
  \epsfxsize=14cm
  \epsffile{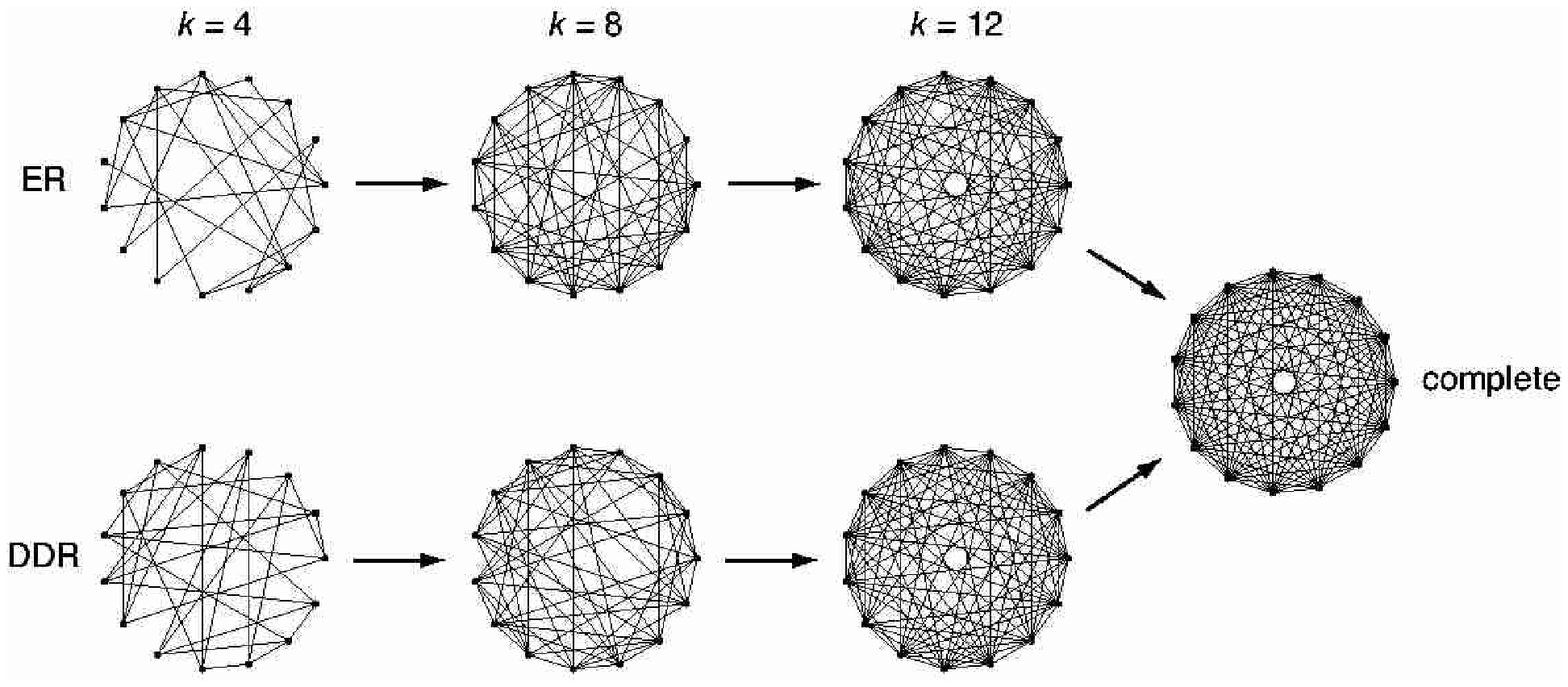}
  \caption{Increase of the degree for \er and \rr
  graphs. We show graphs with 15 nodes for $\nb = 4, 8, 12, 14$. We
  end up with complete graphs for both procedures.} 
  \label{fig:top3_topology}
\end{figure}

We can close the link between the two topological variations by
increasing stepwise the degree $\nb$ in the generated networks, ending
up with complete graphs, as discussed in the beginning of this
section.  In Fig.~\ref{fig:top3_topology} we show the procedures for
small ring lattices for \er and \rr graphs respectively.

\begin{figure}[h]
  \centering
  \epsfxsize=10cm
  \epsffile{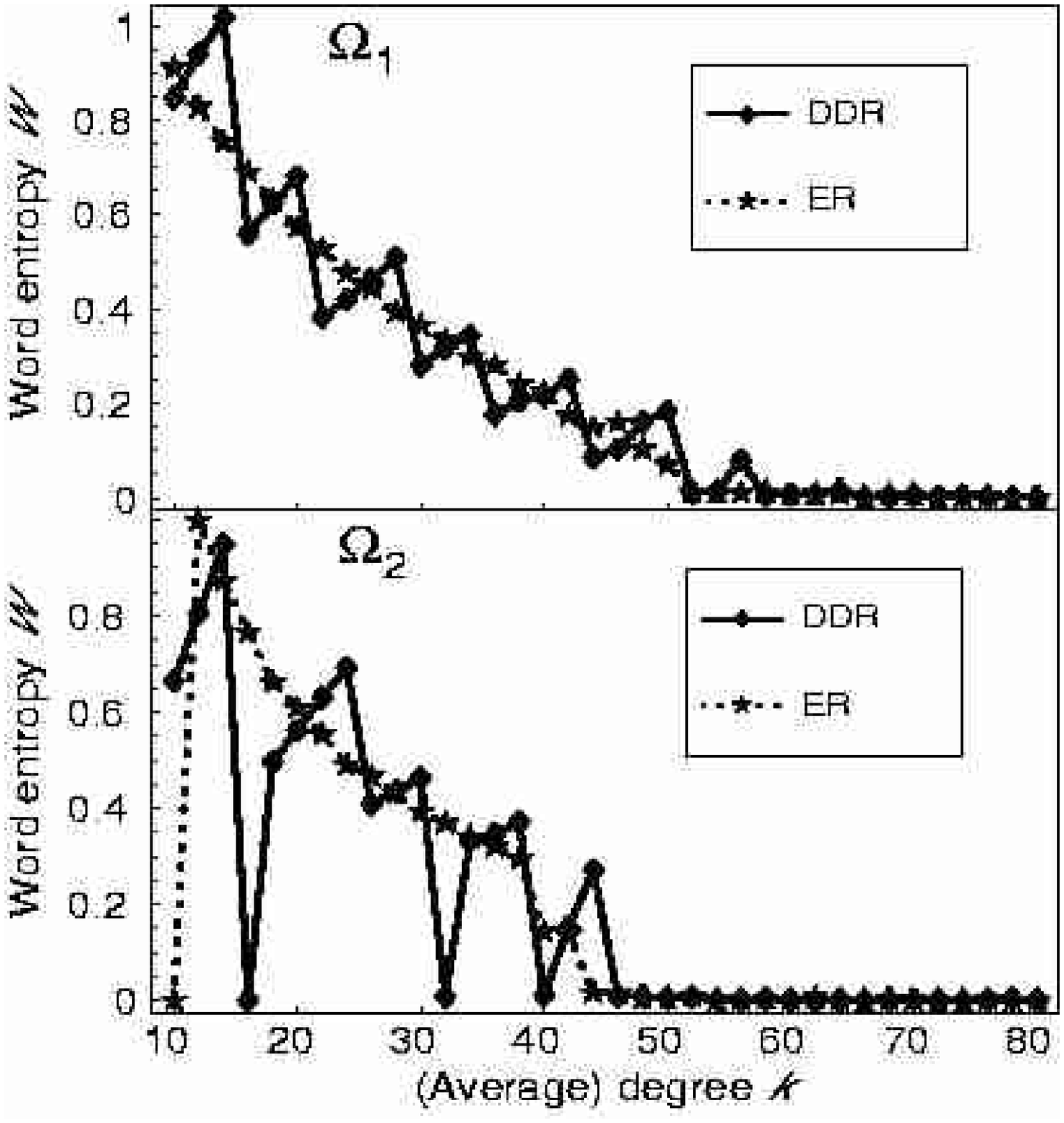}
  \caption{Word entropy $\W$ for increasing $\nb$ for \rr (solid) and
  \er (dotted) graphs and both rule sets. \rr graphs show a peak
  structure with drops to $\W=0$ for \gca2 (bottom). \er graphs show a
  similar overall behavior albeit without peaks and a smoothed curve.
}
  \label{fig:gurke}
\end{figure}

Fig.~\ref{fig:gurke} shows the effect of increasing $\nb$ on the word
entropy $\W$ of patterns of \er and \rr graphs for both $\gca{i}$. The
increase of the neighborhood size results in the leveling of
individual neighborhood density differences. For $\nb > 60$ all nodes
see approximately the same density of 1's within the linked elements
and follow a collective behavior (oscillation for this $\h$). In
contrary to the regular graphs, \er and \rr graphs are both
capable of generating complex patterns for $\gca{2}$.

The class of graphs, which received particular scientific attention in
the last few years, are scale-free graphs \cite{barabasi99} whose
power-law degree distribution is ubiquitous in nature. Therefore, it is
interesting to assess, how topological variations affect the dynamic
behavior and moreover, if these graphs show some kind of extremal
pattern formation capacity.

\begin{figure}[h]
  \centering
  \epsfxsize=10cm
  \epsffile{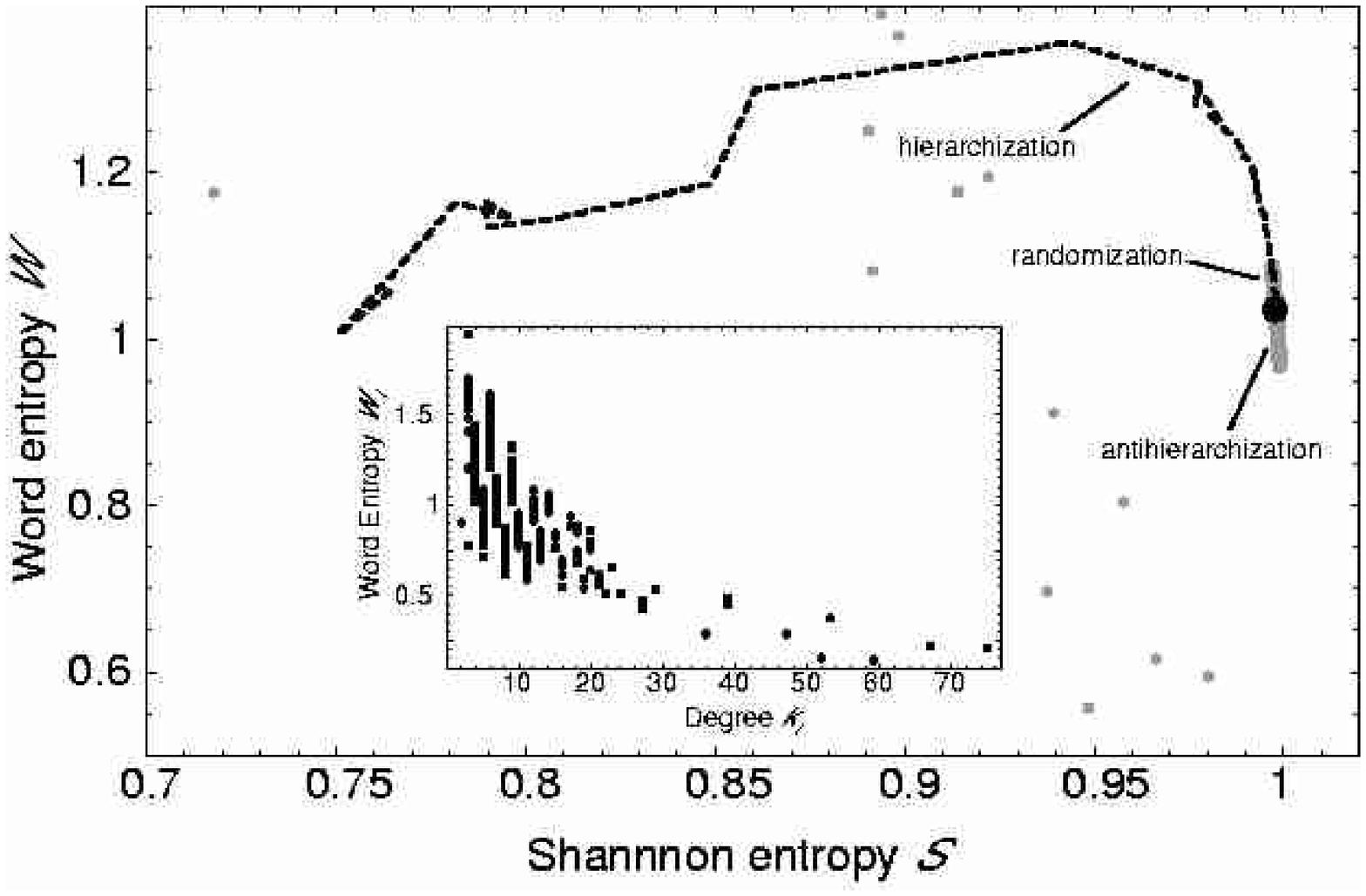}
  \caption{Effects of randomization (dashed), hierarchization (gray) and
    antihierarchization (gray) of a BA scale-free graph with minimal
    degree 4 within the $\ab$ plane. The inset shows the word entropy
    values of 500 single nodes of the graph against their degree.
    Obviously, high $\W_i$ values result from nodes with minimal degree.}
  \label{palme}
\end{figure}

We use the Barabási-Albert (BA) algorithm \cite{barabasi99} to
generate graphs and then apply three algorithms, which change the
topology but keep the degree distribution $P(\nb)$ constant as
described in \cite{trusina04}. We randomize, hierarchize and
antihierarchize the networks by rewiring pairs of edges according to
the corresponding reconnection rule. Randomization rewires two pairs
of linked nodes randomly. The hierarchization process connects nodes
with similar degree and thus imposes an ordering of degrees within the
network. This results in chains of linked cells with increasing degree
and interconnected hubs, i.e., the nodes with extremely high degree
are linked together. Antihierarchization, though, links nodes with
high and low degree and levels hierarchical structures in the network.
The degree correlations of graphs generated with the latter two
procedures resemble the assortative and disassortative mixing observed
in real networks \cite{newman02}.  Fig.~\ref{palme} shows the effect
of randomizing a BA scale-free graph with minimal degree $k_0=4$ up to
a randomization depth of $p=2.5$ within a section of the $\ab$ plane
for $\gca1(0.36)$. Each process affects the response of the system.
While randomization increases and antihierarchization decreases $\W$
slightly at nearly constant $\E$, the hierarchization process yields
maximal $\W$ values for a randomization depth of $p \approx 2$ and a
relatively large variation of $\E$.

The extremely inhomogeneous degree distribution of this kind of graphs
results in an inhomogeneous word entropy distribution, contrary to
regular graphs, where the nodes are indistinguishable with respect to
their degree, and the distribution is delta-like. The inset in
Fig.~\ref{palme} shows $\W_i$ against the degree $\nb_i$ for all nodes
of a BA scale-free graph. Obviously, the hubs in the system account
for small $\W$, while nodes with a minimal degree are responsible for
large and maximal $\W$. Concerning the pattern formation capacity of
hierarchize scale-free graphs, the maximal $\W_i$ generated by single
nodes increases with the degree of hierarchization up to a saturation
level.  This level around $\W=4$ for 2000 hierarchization
steps lies clearly above the maximal values observed for conventional CAs.

\section{Density distribution analysis}
In the previous section we found that a large number of neighbors
synchronizes the system and thus  inhibits complex behavior. In this
section, we want to put the argument on a more quantitative level,
infer the domain of dynamic behavior from the distribution of the
individual densities $P(\rho_i)$ and give an explanation for the
characteristic structure of the word entropy as $\nb$ is increased for
random graphs (Fig.~\ref{fig:gurke}).

The local densities $\rho_i(t)$ and $\tilde{\rho}_i(t)$ defined in
Eqs.~(\ref{eq:r1}) and (\ref{eq:r2}) control the evolution of node $i$ at
time $t$ for both rule sets.  The dynamics therefore depend on the
parameter $\h$ and on the topology of the underlying graph, which
defines the numbers and positions of the linked elements. Let us
consider the case of an initial state density $\rho(0)=0.5$. The
topology then determines the density distribution $P(\rho_i)$. 
In all the following discussions, we assume that $P(\rho_i)$ is time
independent and neglect deformations of $P(\rho_i)$ in the course of
time. In this case, essential features of the dynamics can be immediately derived.
\begin{figure}[h]
  \centering
  \epsfxsize=10cm
  \epsffile{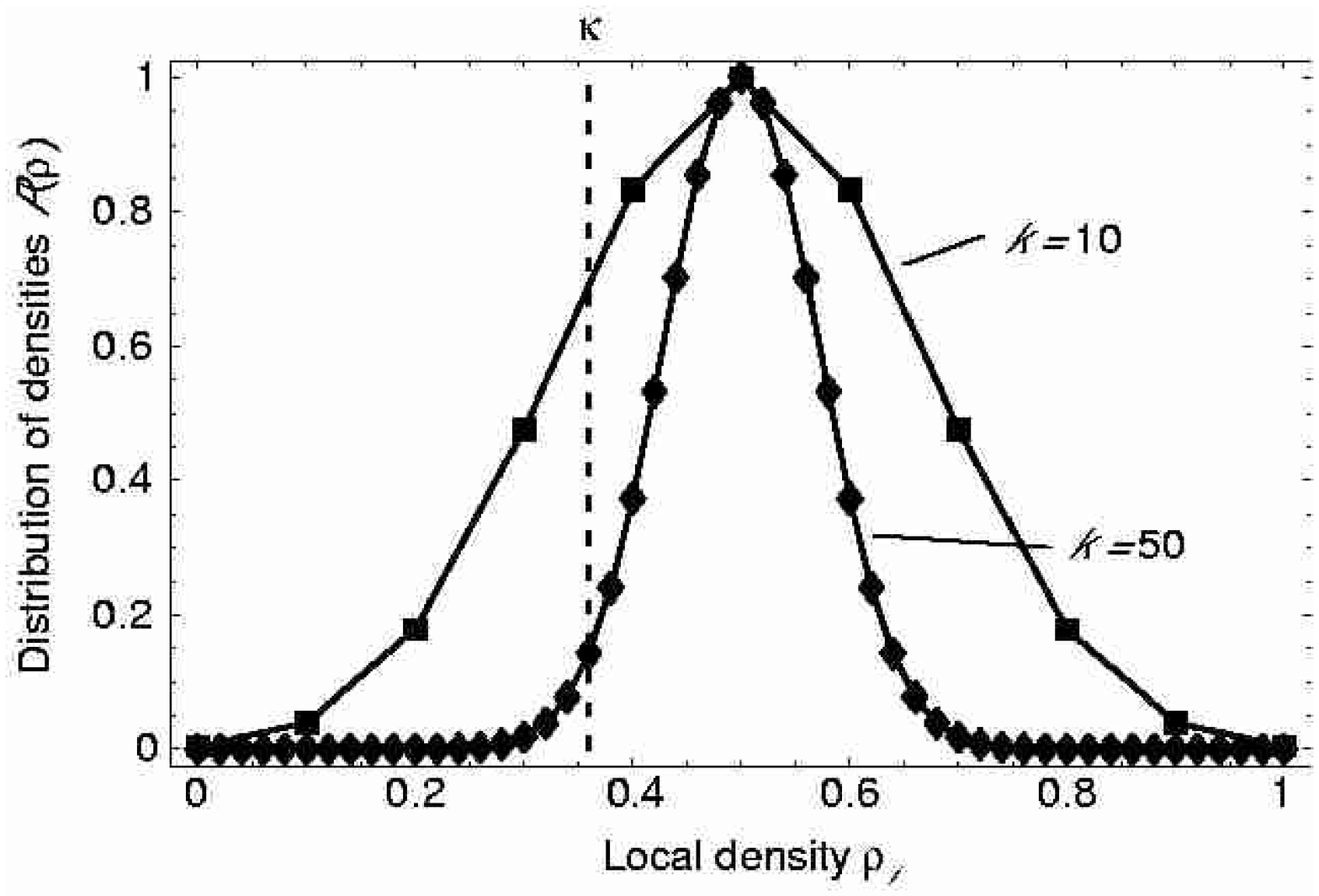}
  \caption{Scaled density distributions $P(\rho_i)$ for \rr graphs
    with $\nb=10$ and $\nb=50$. The line defined by $\h$ indicates the
    ratio of stationary to oscillatory nodes with local densities
    $\rho_i$.}
  \label{fig:risotto}
\end{figure}

In the discussion of Fig.~\ref{fig:sizi}, we argued that for large
$\nb$ all nodes react according to the same density $\rho_i \approx
\rho(0)$ and are therefore synchronized. For smaller $\nb$ values, the
differences in the individual densities $\rho_i$ account for the
coexistence of dynamic behaviors. We can visualize this difference by
plotting the distribution of individual densities, $P(\rho_i)$, for a
small and a large neighborhood, together with the parameter $\h$ in
Fig.~\ref{fig:risotto}. Cells with densities left of the $\h$ line are
stationary, while cells to the right change their state. For a large
neighborhood with $\nb=50$, nearly all nodes oscillate according to
their large local densities $\rho_i>\h$. However, for $\nb=10$, we
find a considerable fraction of cells stationary and argue that this
coexistence of dynamic domains allows for complex behavior.  Note that
the distributions in Fig.~\ref{fig:risotto} are not normalized but
scaled to the identical maximum value. This is done because only the
ratio of oscillatory and stationary cells at fixed $\nb$ provides the
relevant information on the dynamics.

\begin{figure}[h]
  \centering
  \epsfxsize=10cm
  \epsffile{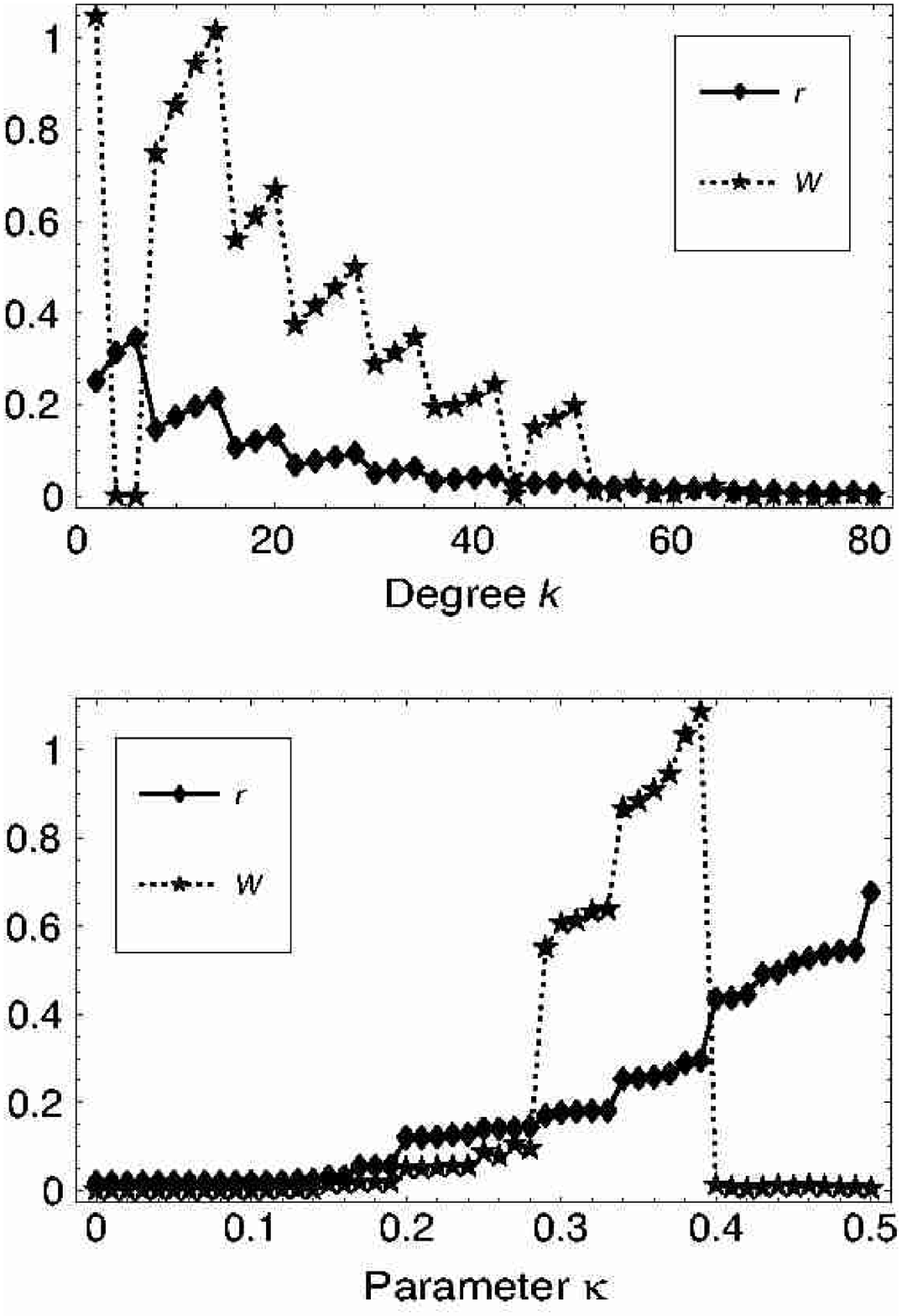}
  \caption{Fraction of constant cells $\constc$ compared with the word
    entropy $\W$ for two random topologies for \gca1. The upper
    picture shows \rr graphs as $\nb$ is increased. The lower one
    shows a BA scale-free graph with minimal degree $\nb_0 = 5$ under
    variation of the parameter $\h$. In both figures a threshold for
    $\constc$ around 0.3 is observable, above which complex patterns
    disappear.}
 \label{fig:yakisoba}
\end{figure}

The peak structure of the word entropy for graphs
with increasing neighborhood size $\nb$ in Figs.~\ref{fig:sizi} and
\ref{fig:gurke} is evident for both regular and \rr graphs. We can
explain this structure for \rr graphs with the following mean-field
approach for $\gca{1}$. Let $\kate_i$ be the
configuration of the coupled states of node $i$, $\kate_i = \{x_{i1},
\ldots, x_{i\nb} \}$. Node $i$ remains constant, if its local
density $\rho_i$ is smaller than $\h$. Thus, the fraction $\constc$ of
constant cells is the sum over all neighborhood configuration with
$\rho \le \h$,
\begin{equation}
  \label{eq:1}
  \constc = \sum_{m=0}^{\lfloor(\h \nb)} p(m, \rho(0)) = \sum_{m=0}^{\lfloor(\h \nb)} \binom{\nb}{m} \rho(0)^m (1-\rho(0))^{\nb
  - m} \; , 
\end{equation}
where $\lfloor(z)$ is the greatest integer less or equal $z$, $m$ the
number of 1's within $\kate_i$ and $p(m)$ is the probability for
finding $m$ out of $k$ neighbors in state 1 for the initial density
$\rho(0)$ and \gca1. In Fig.~\ref{fig:yakisoba}a we plot this number
$\constc$ against the neighborhood size with parameters $\h=0.36$ and
an initial density of $\rho(0)=0.5$ and compare it to the $\W$ values
observed in a numerical simulation. The simulated curve is identical
to the one in Fig.~\ref{fig:gurke}, \gca1, apart from the now enlarged
degree range from 2 to 80.
The direct proportionality between $\constc$ and $\W$ over a wide
range of $\nb$ supports the approach to link the initial settings with
the corresponding dynamic domain. However, a discrepancy for $\nb=4$
and $\nb=6$ is obvious, since the word entropy values are 0 for this
degrees. We argue, that $\constc$ values above $0.3$ inhibit complex
behavior in this \rr graph because of too many stationary
elements. This threshold has to be inferred from the comparison of the
simulated $\W$ and the calculated $\constc$ values.
Nevertheless, together with this threshold, we can predict the
complexity of the time evolution simply from the topology and the
initial local densities for random graphs. This fails for regular
graphs, as can be seen from Fig.~\ref{fig:sizi}, where the shape of
$\W$ differs clearly from the one in Fig.~\ref{fig:yakisoba}. For
those graphs, the density distribution $P(\rho_i)$ is well responsible
for the peak structure and identical at $t=0$ but the spatial
correlations emerging in this topology because of the clustered
neighborhoods alter this distribution in the course of time. 

We can generalize the mean-field approach to other topologies by
incorporating the degree distribution $P(\nb)$ in Eq.~\ref{eq:1}:
\begin{equation}
  \label{eq:g}
  \constc(\nb, \h, \rho(0)) = \sum_{\nb = 0}^{\n-1} P(\nb)
  \sum_{m=0}^{\lfloor(\h \nb')} p(m, \rho(0))  \; 
\end{equation}
In this way we can explain the smoothing of the peak structure for
$\er$ graphs as $\nb$ is increased in Fig.~\ref{fig:gurke}. Moreover,
Eq.~(\ref{eq:g}) can be applied to infer the dynamic behavior of
scale-free graphs. Instead of varying the degree $\nb$, we now study
the dynamic behavior as $\h$ is increased from $0$ to $0.5$.
Fig.~\ref{fig:yakisoba}b shows $\W(\h)$ from a simulation and the
calculated $\constc$ for a BA scale-free graphs with minimum degree
$\nb_0=5$. Again, the qualitative trend and the jumps in the word
complexity are well predicted by $\constc$. For $\constc > 0.3$ the
complex behavior collapses and $\W$ drops sharply to 0. Thus, BA
scale-free graphs can be described properly by the mean-field approach
for random graphs, i.e.\ their local density distribution remains
approximately constant. Notably, with a slight alteration of $p(m,
\rho(0))$, the set of rules \gca2 can be also described by
Eq.~(\ref{eq:g}).

\section{Conclusions}

The role of topology for dynamics is becoming one of the key topics of
nonlinear dynamics and the theory of self-organization.  In this paper
we show by numerical simulation that the pattern formation capacity of
binary CAs strongly depends on the topology of the underlying graph.
We used two temporal entropies, the Shannon entropy $\E$ and the word
entropy $\W$, to separate the different dynamic domains.
We formulated two classes of binary cellular automata on graphs,
$\gca{1}$ and $\gca{2}$, each depending on a single parameter $\h$.
While the first class keeps the Langton parameter $\lambda $ constant,
$\lambda $ varies with $\h$ for the second class.  If applied to
regular graphs, the two sets $\gca{i}$ naturally parameterize and
categorize a subset of CA rules.  Thus, we studied the transition
between stationary and oscillatory behavior for these two CA rule
sets. We assessed the influence of the neighborhood size
$\nb$ for the pattern formation capacity of CAs for a rule that
accounts for increasing $\nb$. For large $\nb$, we observed
synchronization of the elements and the absence of complex patterns.
Beyond the conventional CA topology, we investigated the effect of topological
variations.  We found a crucial dependence on the central element by
comparing the two rule sets. We found that a continuous
change in the topological parameters of a graph can lead to a
continuous trajectory in the $\ab$ plane.
Moreover, we studied the pattern formation capacity of modified BA
scale-free graphs. Hierarchizing such a network leads to a increase in
the word entropy for low-degree nodes. 

The following observations on the link between topology and
  dynamics have been made:
\begin{itemize}
\item The inclusion of the state of cell $i$ itself into the local
  density has a fundamental impact on the emerging dynamics. For all graph
  types we found distinct differences between the two sets $\gca1(\h)$ and
  $\gca2(\h)$ for a large range of the parameter $\h$.
\item Graphs with a delta-like degree distribution display a
  characteristic peak structure in the word entropy as the
  neighborhood size is increased from small values (cf.
  Fig.~\ref{fig:sizi} and Fig.~\ref{fig:gurke} for regular and \rr
  graphs respectively). This crucial dependence on the number of
  neighbors allows for multiple transitions between different dynamic
  domains as $\nb$ grows. This is due to the discrete distribution of
  local densities and can be understood with a mean-field approach.
  However, the local correlations in the CA topology account for the
  enormous jumps in this case.
\item Rewiring a regular topology disintegrates the clustered
  neighborhoods and destroys gradually local collective behavior.
  Beyond the small-world regime, the ability to produce long-range
  correlations is lost. If differences of the rewiring process
  (conservation or alteration of the degree distribution) can be
  resolved or not depends decisively on the definition of the dynamics
  applied (cf. Fig.~\ref{fig:top2_trajectory}).
\item In scale-free graphs a variation of the degree correlations
  changes the word entropy of the time series of individual nodes. The
  maximal $\W_i$ values occur for nodes in the low-degree domains of
  hierarchized topologies and lie far above their regular
  counterparts. However, there are also nodes with small $\W_i$ for all
  degrees, resulting in an average $\W$ within the range of values
  from conventional CAs.
\end{itemize}

Aside from trajectories of topological variations, the path of a graph
along the variation of $\h$ may be used for the topological
characterization. In principle, the trajectory in the $\ab$ plane as a
function of $\h $ (e.g., for $\gca1$) could serve as a dynamical
signature of a particular graph, which assesses the graph's capacity
to display complex dynamics. We found that the two rule sets are
sensitive to variations in different network types. A detailed study
of the connection between the dynamic response and such graph
differences, even up to the motif level, along $\h$, may be an
adequate prerequisite for the characterization of real biological and
technical graphs with the means of the rule sets proposed in this
paper. Eventually, it would be interesting to implement our CA rules,
e.g., on protein interaction graphs and compare this dynamical
signature with those from the graph's randomized counterparts in
search for an evolutionary optimization on this level.

\bibliography{finallit}

\end{document}